\documentclass[a4paper]{article}
\usepackage{epsfig}
\usepackage{graphicx}  
\usepackage{color}     
\usepackage{color}
\usepackage{lscape}
\usepackage{amsmath,amssymb,cite}
\usepackage{hyperref}
\date{\today}
 \normalsize

\def\be {\begin{equation}}
\def\ee {\end{equation}}
\def\bea {\begin{eqnarray}}
\def\eea {\end{eqnarray}}
\def\bc {\begin{center}}
	\def\ec {\end{center}}
\def\bfg {\begin{figure}}
	\def\efg {\end{figure}}
\def\bi {\begin{itemize}}
	\def\ei {\end{itemize}}

\def\no {\noindent}

\def\vs {\vspace}

\def\a  {\alpha}
\def\b  {\beta}

\def\l  {\lambda}

\def\r  {\rho}

\def\t  {\tau}

\def\beq{\begin{equation}}
\def\eeq{\end{equatfion}}
\def\br{\begin{eqnarray}}
\def\er{\end{eqnarray}}
\newcommand{\eel}[1] {\label{#1}\end{equation}}

\newcommand{\bdm}{\begin{displaymath}}
\newcommand{\edm}{\end{displaymath}}
\begin{document}
	\renewcommand{\thefootnote}{\fnsymbol{footnote}}
	
	\vspace{.3cm}

	\title{\Large\bf Green's Function Approach to Entanglement Entropy on Lattices and Fuzzy Spaces}

	\author
	{ {\it \bf    Amel Allouche}  and {\it \bf Djamel Dou\thanks{E.mail: dou-djamel@univ-eloued.dz}}
		\\
		\small Dept of Physics, College of Exact Sciences, Hamma Lakhdar University,
		\\
		\small  El Oued, Algeria.}

	\maketitle

	\begin{center}
		\small{\bf Abstract}\\[3mm]
	\end{center}
	We develop a Green's function  approach to compute R\'{e}nyi entanglement entropy on lattices and fuzzy spaces. The R\'{e}nyi entropy resulting from tracing out an arbitrary collection of subsets of coupled harmonic oscillators  is written as zero temperature partition function generated by an Euclidean action with $n$-fold step potential. The associated Green's function is  explicitly constructed and an alternative new formula for the R\'{e}nyi entropy  is obtained. Finally it is outlined how  this approach can be used to go beyond the gaussian state and include interaction by writing a perturbative expansion for the entanglement entropy .

	\begin{minipage}[h]{14.0cm}
	\end{minipage}
	\vskip 0.3cm \hrule \vskip 0.5cm
	\maketitle
	
	Entanglement is undoubtedly one of the most fundamental concepts that
	distinguishes the quantum world from the classical world. The role that quantum entanglement is playing
	in different areas of physics is impressive. Many measures have been developed to account for the degree of entanglement between different parts of a quantum system \cite{pelino,Nishioka}.  One of the most important and fruitful measures is
	entanglement entropy. Several techniques have been developed to compute this entropy  depending on the area
	of research field \cite{pelino, Nishioka,CC, spinchain, correlation}. In connection with black hole entropy and
	quantum field theory a powerful approach was developed in \cite{W,K} based on the  Euclidean heat kernel
	and Green's function. This approach allows one to make connection with string theory \cite{W,K,suss},
	extract finite  mass dependent but cuttoff independent  contribution to the area law \cite{HW}, and go
	beyond the free case to  investigate the  dependence of entanglement  entropy on the renormalized mass
	\cite{H}.
	
	Although the heat kernel and the Green' function approaches offer  a powerful tool to compute and
	investigate the entanglement entropy they seem to be limited to some special cases of continuum
	geometries.  For example when one traces out half of flat $d+1$ spacetime   R\'{e}nyi entropy turns out
	to be governed by  a simple geometry, namely a cone with deficit angle $2\pi (1-n)$, for which the
	Green's function is known or the heat kernel expansion can be made. However, if one traces out a finite
	region it becomes almost untractable problem to construct the corresponding Green's function or to
	recognize the resulting $n$-sheeted  geometry.
	
	For   field theories on  lattice or on  fuzzy spaces   a Green's
	function or heat kernel approach to entanglement entropy has to our knowledge not yet been developed and the real time formalism is so far the only available technique with its limitation to Guassian states. 
	
	The aim of this work is to  develop  a Green's function approach, a heat kernel as well,   to compute
	entanglement entropy  for a set of $N$-coupled harmonic oscillators (H.Os); for Fermionic osilators we beleive a similar approach could be developed along the same line. Considering $ N$ coupled H.O
	covers indeed a wide class of field theories on lattices and fuzzy spaces due to the fact that many
	physical situations  are modeled by a chain of coupled H.O's ( Fermionic and Bosonic),
	whereas in high energy physics it turns out that for most theories regularized on lattices and fuzzy spaces
	computing the entanglement entropy reduces to considering  independent sectors made  of coupled H.Os, and this is independently of the dimension of the spacetime considered and the geometry of the region traced out  \cite{Nishioka,sd, D1,D2, hol}. More importantly, developing a Green's function technique for field theory on
	lattices and fuzzy spaces would furnish a mathematical framework  on which one can systematically study
	interacting theory using perturbation theory and go beyond the Gaussian state with all associated issues,
	like UV-IR mixing and effect of  mass renormalization, both numerically and analytically using large $N$
	expansion.
	
	Our main result  can be summarized in the following.
	
	Consider a system of $N$-coupled H.O's . The corresponding Largarangian has the standard form
	
	\be
	L= \frac{1}{2}( \dot{Q}^T \dot{Q} -Q^T V Q) ~~~ Q=\left(
	\begin{array}{c}
		q_1 \\
		q_2 \\
		. \\
		. \\
		
		q_N \\
		
	\end{array}
	\right)
	\ee
	
	$V$ is a symmetric positive definite $N \times N$ matrix.
	
	Suppose the whole system is  in the pure ground state $\rho=|0><0|$ and  consider the  density operator $
	\rho_A$ resulting from tracing out  subset $A$, which could be  a union of collection of disjoint subsets, $ \rho_A= \mathrm{Tr}_A \rho$, then we have the following identity

	\be\label{id1}
	\ln \mathrm{Tr} \rho_A^n= -\frac{1}{2} \ln \frac{\det (-\frac{d^2}{d\t^2}+\mathbb{V}(\t))}{\det
		(-\frac{d^2}{d\t^2}+V)^n},~~~~-\infty <\t < \infty 
	\ee
	
	where $ \mathbb{V}(\t)$ is a $n$-fold step-potential given by
	\be
	\mathbb{V} (\t)= \theta(-\t) \mathbb{V}+ \theta(\t) \mathbb{V}_p
	\ee
	
	$\mathbb{V}$ and its image $\mathbb{V}_p$ are $Nn\times Nn $ matrices defined as follows
	\be
	\mathbb{V}= V \otimes \mathbb{I}_{n\times n} ~~  \mathrm{and} ~~  \mathbb{V}_p= \mathbb{P}_\pi
	\mathbb{V}\mathbb{P}_\pi^T
	\ee
	
	 $\mathbb{P}_\pi$ is  $Nn\times Nn $ permutation matrix, its explicit form depends only on the
	choice of the subset $A$ and all we need is its action on $\mathbb{V}$ or the image potential $\mathbb{V}_p$. Each tracing operation is characterized by a permutation matrix  $\mathbb{P}_\pi (A)$, however  different permutation matrices  may lead to the same image potential and therefore to the  same entanglement entropy.   For the important case $A=\{q_{p+1}, q_{p+2},.....q_{N}\}$
	and if we write $V$ as
	\be
	V= \left(
	\begin{array}{cc}
		A & B \\
		B^T & C \\
	\end{array}
	\right)
	\ee
	$A$ and $C$ are respectivelly $(N-p)\times (N-p)$ and $p\times p $ matrices, 
	the matrix $\mathbb{P}_\pi$  leads to the following image potential\footnote{The form of this supermatrix is typical in the calcualtion of the entanglment entropy using Euclidean formalism \cite{W,Jap2}, and here it is understood as arising from the action of a special permutation  matrix.}
	
	\be
	\mathbb{V}_p=
	\left(
	\begin{array}{cccccccc}
		A & 0 & 0 & 0& 0&  \dots  & 0& B \\
		0 & C & B^T &0 & 0 & \dots  & 0& 0 \\
		0 & B &A & 0 & 0 & \dots & 0 & 0\\
		0 & 0 & 0 & C & B^T & \dots & 0& 0\\
		0 & 0 & 0 & B & A & \dots & 0& 0\\
		\vdots & \vdots & \vdots & \vdots & \ddots & \vdots& A& 0 \\
		B^T& 0 & 0 & 0 & \dots & 0 & 0&  C
	\end{array}
	\right)
	\ee
	
	Before proceeding to prove identity (\ref{id1}) we  express $\mathrm{Tr}\r_A^n$  using the Green's
	function associated with $-\frac{d^2}{d\t^2}+\mathbb{V}(\t)$.
	
Using the heat kernel associated with the operators appearing in (\ref{id1}) we can write

	\be\label{heat1}
\ln \mathrm{Tr}\rho_A^n=\frac{1}{2}\int_0^\infty \frac{1}{s}(\mathrm{Tr} \mathbb{ K}_n
(\t,\t,s)-n\mathrm{Tr} K(\t,\t,s)) ds, ~~~~~K(\t,\t,s)= \mathbb{ K}_1(\t,\t,s)
\ee

Define now the Laplace transform of the heat kernel $\mathbb{ K}_n$ , $ \mathbb{ G}_n(\t,\t', E)= \int_{0}^{\infty} e^{-Es} \mathbb{ K}(\t,\t', s) ds$.  The $Nn\times Nn$ matrix function $\mathbb{ G}_n(\t,\t', E)$ is just the Green's function associated with  $-\frac{d^2}{d\t^2}+\mathbb{V}(\t) +E$  satisfying the following differential equation

\begin{equation}\label{GE}
(-\frac{d^2}{d\t^2}+\mathbb{V}(\t)+E)\mathbb{G}_n(\t,\t')=\delta(\t-\t'),~~~~~G(\t,\t')=
\mathbb{G}_1(\t,\t')
\end{equation}

it is then straightforward to show 
\be\label{Gh}
\ln \mathrm{Tr}\rho_A^n=\frac{1}{2}\int_{0}^{\infty} (\mathrm{Tr} \mathbb{ G}_n(\t,\t,E)-n\mathrm{Tr} G(\t,\t,E)) dE  
\ee
The trace $\mathrm{Tr}$ is understood to include integration over $\t$.

The above identity is indeed equivalent  to the evaluation of  $\ln \mathrm{Tr}\rho_A^n$ for massive scalar theory by first evaluating the derivative of  $\ln  \mathbb{Z}_n$   with respect to $m^2$, with the role of $m^2$ is here played  by $E$,  expressing it in terms of the coincident Green's function $\mathbb{ G}_n(\t,\t,m^2)$  on the $n$-fold cover space and then integrating with respect to $m^2$   \cite{CC}. 
	
	It remains now to find the explicit form of $\mathbb{G}_n(\t,\t')$. In view of the fact that $ \mathbb{V}$
	and $\mathbb{V}_p$ do not generally commute, the construction of the Green's matrix is not straightforward. Actually it is the
	nonvanishing of the commutator $[ \mathbb{V}, \mathbb{P}_\pi] $  which will
	lead to non zero entanglement entropy, therefore $ \|[ \mathbb{V}, \mathbb{P}_\pi] \|/\|\mathbb{V}\| $ can be considered in a 
	sense a measure of the degree of entanglement\footnote{The matrix norm $\|.\|$ could be the Hilbert-Schmidt norm or whatever matrix norm one prefers.}.
	
	To construct $\mathbb{G}_n(\t,\t')$ we followed the method  developed in \cite{meftah} and wrote down a
	perturbation series for it and used Laplace transform method to turn it into Weiner-Hopf problem, then
	solving the perturbation series order by order  the exact form of $\mathbb{G}(\t,\t')$ could be constructed.
	In this letter we  only give the final result and the detail will be reported elsewhere  \cite{AD}.
	
	\be\label{GFS}
	\mathbb{G}_{\mp\mp}(\t,\t') = \frac{1}{2\mathbb{W}_\mp} e^{-\mathbb{W}_\mp|\t-\t'|}
	+\frac{1}{2\mathbb{W}_\mp}  e^{\pm\mathbb{W}_\mp\t} \bigg[
	\frac{\mathbb{W}_\mp-\mathbb{W}_\pm}{\mathbb{W}_-+\mathbb{W}_+}  \bigg] e^{\pm\mathbb{W}_\mp \t'}
	\ee
	
	and

	\be
	\mathbb{G}_{\pm\mp}(\t,\t')=  e^{\mp\mathbb{W}_\pm\t}\frac{1}{\mathbb{W}_-+\mathbb{W}_+ }
	e^{\pm\mathbb{W}_\mp \t'}
	\ee
	
	where we used the standard notation $\mathbb{G}_{+-}(\t,\t')=\mathbb{G}_n(\t,\t') $ for $\t >0, \t'<0$
	..etc and with
	$$
	\mathbb{W}_{\pm}= \sqrt{\mathbb{V}_\pm},~~\mathbb{V}_- = \mathbb{V}+E ,~~~~~\mathbb{V}_+ \mathbb{=V}_p
	+E
	$$
	
	$$
	\frac{1}{{W}_{\pm}}={W}_{\pm}^{-1},~~ \mathrm{and} ~~
	\frac{\mathbb{W}_\mp-\mathbb{W}_\pm}{\mathbb{W}_-+\mathbb{W}_+} \equiv
	(\mathbb{W}_\mp-\mathbb{W}_\pm)(\mathbb{W}_-+\mathbb{W}_+)^{-1}
	$$
	
	It is not difficult to check that above constructed $\mathbb{G}(\t,\t')$ is a solution of eqt (\ref{GE})
	and satisfies along with its first derivatives all the continuity conditions at the cut. When $[ \mathbb{V}, \mathbb{V}_p]=0$ or $\mathbb{V}=0, \mathbb{V}_p= V $, $V$ a real number and with  the appropriate redefinitions of the parameters,  $\mathbb{G}_n(\t,\t')$  reduces to the Green's function for a one dimensional  step potential \cite{meftah,stepp}.

	Substituting  $\mathbb{G}_n(\t,\t)$ into eqt (\ref{Gh}) and performing the integration with respect to $\t$ we obtain

	\be\label{ent3}
	\ln \mathrm{Tr}\rho_A^n=  \frac{1}{8}\int_{0}^{\infty} \mathrm{Tr}  (
	\mathbb{V}_-^{-1}-\mathbb{V}_+^{-1})(\mathbb{W}_--\mathbb{W}_+)\mathbb{(W}_-+\mathbb{W}_+)^{-1} dE
	\ee

	Equation (\ref{ent3}) is  our   alternative formula which gives R\'{e}nyi entropy  resulting from tracing out an arbitrary subset of H.Os, and it can be used  to compute the
	entanglement entropy via the replica trick. The expression of $\ln \mathrm{Tr}\rho_A^n$  is invariant under  rescaling of the potential and vanishes when $ [ \mathbb{V}, \mathbb{P}_\pi]=0$ as it should be.
	
 We now return to the proof of our starting identity
	(\ref{id1}). The proof of (\ref{id1}) is quite easy and makes   use of the Euclidean path integral to express $ Tr\rho_A^n$.
	
	We illustrate the proof for the case where we trace out a subset $A=\{q_{p+1}, q_{p+2},.....q_{N}\}$, the general case is similar. Let
	$\bar{A}$ denote the complimentary subset, $Q_A\equiv(q_{p+1}, q_{p+2},.....q_{N})$  and
	$Q_{\bar{A}}\equiv(q_{1}, q_{1},.....q_{p})$.
	
	The ground state density operator can be written using the path integral as
	
	\be
	\r (Q'_{\bar{A}},Q'_{A} ;Q''_{\bar{A}},Q''_A) =\int \mathcal{D}Q \exp
	(-\int_{-\infty(Q=0)}^{0^-(Q=(Q'_{\bar{A}},Q_{A}'))}L_E
	d\t-\int_{0^+(Q=(Q''_{\bar{A}},Q''_{A}))}^{\infty(Q=0 )}L_Ed\t)
	\ee
	
	The reduced density matrix is then given by,
	\be
	\r_A (Q'_{\bar{A}} ;Q''_{\bar{A}})=\int  \r (Q'_{\bar{A}},Q_{A}
	;Q''_{\bar{A}},Q_A)dQ_A,~~~~~dQ_A=\prod_{A}dq_i
	\ee
	from which it follows that
	
	\be
	\mathrm{Tr}_{\bar{A}}\r_A^n=\int \r_A (Q^1_{\bar{A}} ;Q^2_{\bar{A}})\r_A (Q^2_{\bar{A}}
	;Q^3_{\bar{A}})\cdot \cdot \cdot \cdot\r_A (Q^{n}_{\bar{A}} ;Q^1_{\bar{A}}) dQ^1_{\bar{A}}
	dQ^2_{\bar{A}}\cdot\cdot\cdot dQ^{n}_{\bar{A}}
	\ee
	
	Now if we define an $n$-fold field variable $\mathbb{Q}$ with $Nn$ components as
	\be
	\mathbb{Q}= \left(
	\begin{array}{c}
		Q^1_{\bar{A}} \\
		Q^1_A \\
		Q^2_{\bar{A}}  \\
		Q^2_A \\
		\cdot\\
		\cdot\\
		\cdot\\
		
		Q^n_{\bar{A}} \\
		Q^n_{A}
		
	\end{array}
	\right)
	~~   \mathrm{and}~~~\mathbb{V}= V \otimes
	\mathbb{I}_{n\times n}
	\ee
	
	then it is not difficult to show that
	\be
	\mathrm{Tr}_{\bar{A}}\r_A^n =\int \mathcal{D}\mathbb{Q} \exp
	(-\int_{-\infty,\mathbb{Q}=0}^{0^-,\mathbb{Q}=(Q^1_{\bar{A}},Q^1_{A};Q^2_{\bar{A}},Q^2_{A}\cdot\cdot\cdot
		Q^n_{\bar{A}},Q^n_{A} )}\mathbb{L}_E
	d\t-\int_{0^+,\mathbb{Q}=(Q^2_{\bar{A}},Q^1_{A};Q^3_{\bar{A}},Q^1_{A}\cdot\cdot\cdot
		Q^n_{\bar{A}},Q^{n-1}_{A};Q^1_{\bar{A}},Q^{n}_{A})}^{\infty, \mathbb{Q}=0}\mathbb{L}_E d\t)
	\ee
	
	where $\mathbb{L}_E$ is given by
	\be
	\mathbb{L}_E=\frac{1}{2}( \dot{\mathbb{Q}}^T \dot{\mathbb{Q}} +\mathbb{Q}^T\mathbb{ V} \mathbb{Q})
	\ee
	What prevents us from writing $\mathrm{Tr}_{\bar{A}}\r_A^n$ as a zero temperature partition function with
	specified Lagrangian are the boundary conditions on the variables $\mathbb{Q}_{\bar{A}}$ at the cut. This
	obstacle can be overcome by moving the cut or the discontinuity to the potential by introducing a
	permutation matrix $ \mathbb{P}_\pi$ which maps
	$(Q^1_{\bar{A}},Q^1_{A};Q^2_{\bar{A}},Q^2_{A}\cdot\cdot\cdot Q^n_{\bar{A}},Q^n_{A} )$ into
	$(Q^2_{\bar{A}},Q^1_{A};Q^3_{\bar{A}},Q^2_{A};\cdot\cdot\cdot
	Q^n_{\bar{A}},Q^{n-1}_{A};Q^1_{\bar{A}},Q^{n}_{A})$. Then $ \mathrm{Tr}_{\bar{A}}\r_A^n$  with the correct normatizalion can be written as
	\be\label{partionf}
	\mathrm{Tr}_{\bar{A}}\r_A^n = \frac{\mathbb{Z}_n}{\mathbb{Z}_1^n}=\frac{\int \mathcal{D} \mathbb{Q}
		e^{-\mathbb{S}_E}}{(\int \mathcal{D}Q e^{-S_E})^n}
	\ee
	where
	\be\label{action}
	\mathbb{S}_E=\int_{-\infty}^{\infty} ( \frac{1}{2}[ \dot{\mathbb{Q}}^T \dot{\mathbb{Q}}
	+\mathbb{Q}^T\mathbb{ V(\t)} \mathbb{Q}] d\t, ~~~\mathbb{V} (\t)= \theta(-\t) \mathbb{V}+ \theta(\t)
	\mathbb{V}_p
	\ee
	
	then indentity (\ref{id1}) follows immediately.
	
	This completes the proof of our starting identity.
	
	Let us now see  equation (\ref{ent3}) at work by using it to recover the standard formula for the entanglement entropy for $2$ coupled
	H.O's \cite{sd, sor}. This turned out be a straightforward exercise and only the main steps are outlined.  For realistic models exact results are generally impossible, however we believe
	that formula (\ref{ent3}) will be very suitable for using Discrete Fourier Transform and large $N$ expansion.
	
	We consider the following Hamiltonian \cite{sd} $H=\frac{1}{2}[p_1^2+p_2^2 +k_0(q_1^2+q_2^2)+k_1(q_1-q_2)^2]$, then the relevant corresponding potential is of the form\footnote{This form can always be achieved by rescaling the potential without affecting the entropy.} $V=\left(
	\begin{array}{cc}
	a & 1 \\
	1 & a \\
	\end{array}
	\right)$

 It is easy to show that
	\be\label{c1}
	(\mathbb{V}_-^{-1}-\mathbb{V}_+^{-1})(\mathbb{W}_--\mathbb{W}_+) =\frac{(\sqrt{x-1}-\sqrt{x+1})}{2(x^2-1)} \mathbb{C}
	\ee
	where $x=a+E$, $ \mathbb{C}$ is a $2n\times 2n$ circulant matrix with elements
	$ c_j$, $j=0,1.\cdot \cdot\cdot 2n-1$, $c_0=2, c_{2n-2}=c_2=-1...$  and zeros otherwise.  On the other hand $ \mathbb{W}_-+\mathbb{W}_+$  is in this particular case another circulant matrix   $\mathbb{C}'$ with elements   $c'_0=2 \a_+, c'_{2n-1}=c'_1=\a_-, \a_{\pm}(x) = \sqrt{x+1}\pm \sqrt{x-1}$ and zeros otherwise.
	Because circulant matrices form a commutative subalgebra $\mathbb{C}$ and $\mathbb{C}'^{-1}$
	can  simultaneously be diagonalized. The eigenvalues of $\mathbb{C}$ and $\mathbb{C}'$ are
	given by
	
	$$
	\l_j =2(1-\cos 2\frac{\pi j}{n}),~~~\l'_j= \a_++\a_- \cos\frac{ \pi j}{n}
	$$
	then it follows  from (\ref{ent3}) that
	\be\label{ent2HO}
	\ln \mathrm{Tr}\rho_A^n= -\frac{1}{8}\sum_{j=0}^{2n-1} (1-\cos \frac{2\pi j}{n})\int_{a}^{\infty} \frac{\a_-(x)}{(x^2-1)(\a_++\a_-\cos\frac{\pi j}{n})} dx
	\ee
	The integrals inside the sum can be exactly evaluated and after some arrangements and  simplifications we end up with
	
	\be\label{sum}
	\ln \mathrm{Tr}\rho_A^n= -\sum_{j=1}^{n-1} \ln  \frac{cos \frac{\pi j}{n}+\b }{\sqrt{\b^2-1}} 
	\ee
	
	where $\b=a+\sqrt{a^2-1}$.
	
	The sum  in (\ref{sum} ) can be evaluated exactly using standard products and sum formulas leading to
	\be\label{reyni}
	\ln \mathrm{Tr}\rho_A^n= \ln \frac{(\a^2-1)^{n}}{(\a^{2n}-1)}, ~~~~~\a=\b+\sqrt{\b^2-1}
	\ee
	
	Here our $\alpha$ turns out to be related to $\xi$ defined in \cite{sd} by $\a=\frac{1}{\sqrt{\xi}}$.
	
	Using the replick trick and the definition of $\a$ it is straightforward to show that
	\be
	-\frac{d}{dn}\ln \mathrm{Tr}\rho_A^n|_{n=1}= -\ln (1-\xi) -\frac{\xi}{1-\xi}\ln \xi
	\ee

	Finally we turn our attention to an important application of the present approach, namely  interacting theories. Here we just outline how the leading order correction  to entanglement entropy  can be included, higher order and applications to interacting scalar theories on  the fuzzy sphere and Moyal plan will be considered in \cite{AD}.
	
	Let us consider $\lambda \phi^4$ type interaction. The lagrangian is of the form\footnote{Here we are not having in mind a naive lattice  discretization for $\phi^4$ theory, otherwise we would write the interaction as $\sum q_i^4$ and the form of the correction would be slightly different for the one given above and more similar to the expansion given in \cite{H} in the continuum.}
	
	\be\label{lagint}
	L_E=\frac{1}{2}( \dot{Q}^T \dot{Q} +Q^T V Q)+\frac{\lambda}{4} (Q^TQ)^2
	\ee
	In view of the fact that the interaction term is invariant under the action of the perumtation matrix the perturbation series for $\ln \mathbb{Z}_n(\lambda)$ is straightforward and formally similar  to the calculation of \cite{H} in the continuum.
	
	\be\label{enper}
	\ln \mathrm{Tr}\rho_A^n(\lambda)= \ln \mathbb{Z}_n(\lambda)-n\ln Z_1(\lambda)
	\ee
	
	expanding to the leading order in $\lambda$ and using Wick's theorem we obtain\footnote{The Green's function  is  understood here to be avaluated for $E=0$.}
	
	\be\label{enper1}
	\ln \mathrm{Tr}\rho_A^n(\lambda)= \ln \mathbb{Z}_n(0)-n\ln Z_1(0)-\frac{\l}{2}\int d\t \big[ \mathrm{Tr} \mathbb{G}(\t,\t)^2 -n \mathrm{Tr} G(\t,\t)^2\big ]+ \mathcal{O}(\l^2)
	\ee
	
	using the expression of $\mathbb{G}(\t,\t)$ one can show that the leading order correction is formally given by
	\be\label{foc}
	\delta \ln \mathrm{Tr}\rho_A^n= -\frac{\lambda}{2} \bigg[ -\frac{3n}{8} \mathrm{Tr}V^{-3/2}+\frac{1}{4} \mathrm{Tr}(\mathbb{V}_+^{-1}+\mathbb{V}_-^{-1})(\mathbb{W}_-+\mathbb{W}_+)^{-1} +F(\mathbb{W}_-,\mathbb{W}_+)+F(\mathbb{W}_+,\mathbb{W}_-) \bigg]
	\ee
	where $F $ is defined by
	$$
	F(\mathbb{W}_-,\mathbb{W}_+)= \int_{0}^{\infty} \mathrm{Tr}\big[ (\mathbb{W}_-+\mathbb{W}_+)^{-1} e^{-2\mathbb{W}_+\t} (\mathbb{W}_-+\mathbb{W}_+)^{-1}e^{-2\mathbb{W}_+\t}\big]
	$$
	
As can be seen  at this leading order the tunnelling parts of the Green's function $\mathbb{G}_{\pm\mp}$ still plays no role, however   they start to contribute at the next  order.

We conclude by some  remarks about the present approach . For a fixed $n >1$ formulas (\ref{ent3}) and (\ref{enper1}) can be used  directly to compute  numerically R\'{e}nyi entropy  and include corrections due to interaction, however to carry out analytical study or to consider  the important case $n=1$   analytical techniques must be used or developed. Generally the complexity of the study depends on the nature of the region $A$ to be traced out which in turn changes the permutation matrix $\mathbb{P}_\pi$ leading to different image matrices, $\mathbb{V}_p, \mathbb{W}_p$, nevertheless  it turns out  that for many cases of interest Toeplitz and Block Toeplitz  matrices  play an essential role within the present approach. For instance it turned out that for scalar theories on fuzzy spaces the dominant contribution to the entanglement entropy comes from regions where the potential is essentially a Toeplitz matrix \cite{D2}, whereas for $1+1$ scalar theory on lattice the problem could be reduced to computing the trace over product of Toeplitz matrices \cite{AD},  and we hope that this will make many problems analytically tractable in the large $N$ limit . Actually large $N$ expansion in this approach should be understood as a discrete version of the heat kernel expansion, therefore we expect the mathematical tools and theorems developed for the asymtotic behavior of matrices, in particular for Toeplitz matrices \cite{Top}, will be indispensable in order to obtain anlytical results.
	
	\vs{.2cm}
	\no {\bf Acknowledgment}
	\no
	
	We thank P. Calabrese for useful discussions.
	
	D.Dou would like to thank the High Energy section of ICTP for their kind hospitality.
	
	This work is supported in part by the {\it Lab of Linear Operators and Partial Diff Equations, Theory
		and Application, Hamma Lakhdar University, El Oued , Algeria}.
	%
	

\end{document}